\documentclass[aps,amsfonts,reprint,tightenlines,amssymb,superscriptaddress,twocolumn,floatfix]{revtex4-2}  

\usepackage{graphicx}
\usepackage{dcolumn}
\usepackage{bm}
\usepackage{mathtools}
\usepackage{amsmath}
\usepackage{amsthm}
\usepackage[utf8]{inputenc}

\usepackage{qcircuit}
\usepackage[caption=false]{subfig}

\usepackage{algorithm}
\usepackage[noend]{algpseudocode}
\algrenewcommand\algorithmicdo{}

\makeatletter
\renewcommand{\ALG@name}{Procedure}
\makeatother

\usepackage[linktocpage=true,
  colorlinks=true, 
  pdfborder={0 0 0},
  linkcolor=blue,
  citecolor=red,
  filecolor=yellow,
  urlcolor=blue,
  bookmarks,
  pdfauthor={},
]{hyperref}
\usepackage{orcidlink}

\usepackage{ifthen}
\newcounter{is_qcircuit_used}
\newcommand{\bs}{\boldsymbol}

\begin{document}

\preprint{APS/123-QED}

\title{Orbital-free density functional theory with first-quantized quantum subroutines}

\author{Yusuke Nishiya\orcidlink{0000-0001-6526-0936}}
\email{ynishiya@quemix.com}
\affiliation{
  Department of Physics, 
  Graduate School of Science, 
  The University of Tokyo,
  7-3-1, Hongo, Bunkyo-ku, Tokyo 113-0033, Japan
}
\affiliation{
Quemix Inc.,
Taiyo Life Nihombashi Building,
2-11-2,
Nihombashi Chuo-ku, 
Tokyo 103-0027,
Japan
}

\author{Hirofumi Nishi\orcidlink{0000-0001-5155-6605}}
\affiliation{
  Department of Physics, 
  Graduate School of Science, 
  The University of Tokyo,
  7-3-1, Hongo, Bunkyo-ku, Tokyo 113-0033, Japan
}
\affiliation{
Quemix Inc.,
Taiyo Life Nihombashi Building,
2-11-2,
Nihombashi Chuo-ku, 
Tokyo 103-0027,
Japan
}

\author{Taichi Kosugi\orcidlink{0000-0003-3379-3361}}
\affiliation{
 Department of Physics, 
 Graduate School of Science, 
 The University of Tokyo,
 7-3-1, Hongo, Bunkyo-ku, Tokyo 113-0033, Japan
}

\affiliation{
Quemix Inc.,
Taiyo Life Nihombashi Building,
2-11-2,
Nihombashi Chuo-ku, 
Tokyo 103-0027,
Japan
}

\author{Yu-ichiro Matsushita\orcidlink{0000-0002-9254-5918}}
\affiliation{
  Department of Physics, 
  Graduate School of Science, 
  The University of Tokyo,
  7-3-1, Hongo, Bunkyo-ku, Tokyo 113-0033, Japan
}
\affiliation{
Quemix Inc.,
Taiyo Life Nihombashi Building,
2-11-2,
Nihombashi Chuo-ku, 
Tokyo 103-0027,
Japan
}
\affiliation{
Quantum Material and Applications Research Center,
National Institutes for Quantum Science and Technology (QST),
2-12-1, Ookayama, Meguro-ku, Tokyo 152-8552, Japan
}
\affiliation{
Laboratory for Materials and Structures,
Institute of Innovative Research,
Tokyo Institute of Technology,
Yokohama 226-8503,
Japan
}

\date{\today}

\begin{abstract}
In this study, we propose a quantum-classical hybrid scheme for performing orbital-free density functional theory (OFDFT) using probabilistic imaginary-time evolution (PITE), designed for the era of fault-tolerant quantum computers (FTQC), as a material calculation method for large-scale systems. 
PITE is applied to the part of OFDFT that searches the ground state of the Hamiltonian in each self-consistent field (SCF) iteration, while the other parts such as electron density and Hamiltonian updates are performed by existing algorithms on classical computers. When the simulation cell is discretized into $N_\mathrm{g}$ grid points, combined with quantum phase estimation (QPE), it is shown that  obtaining the ground state energy of Hamiltonian requires a circuit depth of $O(\log N_\mathrm{g})$. The ground state calculation part in OFDFT is expected to be accelerated, for example, by creating an appropriate preconditioner from the estimated ground state energy for the locally optimal block preconditioned conjugate gradient (LOBPCG) method.
\end{abstract}

\maketitle 

\section{Introduction}
Materials computation is one of the most promising applications of quantum computation.
Determining the ground state of electrons in materials and performing Hamiltonian simulations with high speed and accuracy will accelerate drug discovery and the new materials search, leading to a paradigm shift toward higher efficiency and energy conservation in all areas of industry.
Previously proposed quantum algorithms for ground state calculations include the variational quantum eigensolver (VQE)~\cite{Peruzzo2014Ncom, Farhi2014}, adiabatic time evolution~\cite{Farhi2000arXiv, Aspuru-Guzik2005Science, Nishiya}, and imaginary time evolution (ITE) on a quantum computer~\cite{gingrich2004non, terashima2005nonunitary, Yuan2019Quantum, Mcardle2019npjQI, Motta2020NPhys, Benedetti2021PRR, Kosugi2022PhysRevResearch, Leadbeater2024QST}. Almost all of them encode the many-electron wavefunction into the state of a quantum register consisting of many qubits to obtain the exact ground state of the {\it ab initio} Hamiltonian.

On the other hand, density functional theory (DFT)~\cite{HohenbergKohn1964} has been widely used and succeeded in computation of many-electron systems using classical computers.
Although DFT is less accurate than approaches that deal directly with many-electron wave functions, it often provides sufficient accuracy to describe physical phenomena. Furthermore, it is the preferred choice for calculations across a wide range of materials due to its significantly lower computational cost.
In particular, orbital-free DFT (OFDFT), a type of DFT in which the kinetic energy functional is explicitly given as a functional of the electron density, can be performed at a computational cost of $O(N)$ to $O(N \log N)$ for $N$ atoms~\cite{delRioCarter_review2018, Mi_review2023}, and is expected to be applied to larger systems.

Materials calculations using quantum algorithms should not only pursue accuracy, but should also be able to handle large-scale systems. 
In particular, it is desirable to be able to target a system size that cannot be handled by current classical computers, and ultimately, to be able to simulate polymers such as proteins, amorphous materials, surfaces and interfaces of materials containing line and planar defects without any coarse-graining.
In this study, we propose a quantum-classical hybrid scheme of OFDFT as a first step in the development of quantum algorithms for large-scale materials computations. The proposed method includes probabilistic imaginary-time evolution (PITE)~\cite{Kosugi2022PhysRevResearch, Kosugi2023npjqi} in the first quantization form and quantum phase estimation (QPE)~\cite{Kitaev1995arXiv, Abrams1999PRL, Wiebe2016PRL, OBrien_2019NJP, Yamamoto2024PRR, Ding2023PRXQ} as subroutines for the ground-state calculation of a given Hamiltonian.

\section{Methods}
\subsection{Orbital-free density fucntional theory}
DFT~\cite{HohenbergKohn1964} states that the total energy of an electron system can be expressed as a functional of the electron density $\rho(\bs{r})$ alone. The electron density that minimizes this functional corresponds to the ground state electron density. In this case, the total energy functional is constructed as follows:
\begin{gather}
E[\rho] = T_S[\rho] + \int v_\text{ext}(\boldsymbol{r})\rho(\bs{r}) d\boldsymbol{r} + E_\text{H}[\rho] + E_\text{XC}[\rho],
\label{eq:etot}
\end{gather}
where $T_S$ is the kinetic energy of the non-interacting system, $E_\mathrm{H}$ is Hartree energy that represents the classical electron-electron interaction, and $E_\mathrm{XC}$ is the exchange-correlation energy.
In this study, we assume the use of local pseudopotentials, such as bulk-derived local
pseudopotential (BLPS)~\cite{Huang2008, HuangCarter2010PRB, Zhou2004PRB, delRio2017} and {\it ab initio} local ionic pseudopotential (LIPS)~\cite{Imoto2021PRR}, as the potential $v_\mathrm{ext}$ representing the interaction between valence electrons and nuclei with core electrons.
In OFDFT, the kinetic energy functional is directly given as a functional of the density, and optimization of the electron density can be translated into finding the eigenvector corresponding to the smallest eigenvalue of the following self-consistent eigenvalue problem~\cite{Levy1984PRA, Imoto2021PRR}:
\begin{gather}
\hat{\mathcal{H}}_\text{OF}^{(\lambda)}[\rho]\sqrt{\rho(\bs{r})} = \frac{\mu}{\lambda}\sqrt{\rho(\bs{r})},
\label{eq:ofdft}
\end{gather}
where
\begin{gather}
\hat{\mathcal{H}}_\text{OF}^{(\lambda)}[\rho] \equiv -\frac{1}{2} \nabla^2 + \frac{1}{\lambda}\{v_\text{KS}([\rho];\bs{r}) + v_\text{r}([\rho];\bs{r})\},
\label{eq:of_hamiltonian}
\end{gather}
\begin{gather}
v_\text{KS}([\rho];\boldsymbol{r}) \equiv v_\text{ext}(\bs{r}) + \int \frac{\rho(\bs{r})'}{|\bs{r}-\bs{r}'|} d\bs{r}' + \frac{\delta E_\text{XC}}{\delta \rho}(\bs{r}),
\end{gather}
\begin{gather}
v_\text{r}([\rho];\boldsymbol{r}) \equiv \frac{\delta T_S}{\delta \rho}(\bs{r}) - \lambda \frac{\delta T_\text{vW}}{\delta \rho}(\bs{r}).
\end{gather}
$v_\text{r}$ is defined by $T_S$ in Eq.~(\ref{eq:etot}) and von-Weitzs\"{a}cker functional $T_\text{vW}$~\cite{vWKEDF}, which satisfies $\delta T_\text{vW}/\delta \rho = - (\nabla^2 \sqrt{\rho}) /(2 \sqrt{\rho})$. 
$\lambda$ is a constant that can be set arbitrarily by the user to improve the convergence of the calculation.
Eq.~(\ref{eq:ofdft}) is typically solved by the following procedure.
First, we find the ground state $\sqrt{\rho_\text{out}}$ of $\hat{\mathcal{H}}_\text{OF}^{(\lambda)}[\rho_\text{in}]$ defined by a certain input electron density $\rho_\text{in}$.
Then, from the obtained $\rho_\text{out}$ or their history, the next $\rho_\text{in}$ is determined and the Hamiltonian $\hat{\mathcal{H}}_\text{OF}^{(\lambda)}[\rho_\text{in}]$ is updated.
The above procedure is repeated, and the calculation is completed when $\| \rho_\text{in} - \rho_\text{out}\| < \varepsilon$ with convergence threshold $\varepsilon$.
This iterative density update to reach the self-consistent solution is referred to often as self-consistent field (SCF) iteration.

\subsection{Probabilistic imaginary-time evolution method}
We outline PITE, which is a ground-state calculation method for the Hamiltonian $\hat{\mathcal{H}}$ of an $n$-qubit system\cite{Kosugi2022PhysRevResearch}.
Let us consider the non-unitary and Hermitian operator $\mathcal{M} = m_0 e^{-\hat{\mathcal{H}}\Delta \tau}$, where $m_0$ is an adjustable real parameter that satisfies $0<m<1$ and $m_0 \neq 1 /\sqrt{2}$, and is introduced to avoid singularities that appear when performing the Taylor expansion to derive the approximate PITE described below.
The unitary operation acting on the $n+1$-qubit system is defined by introducing one ancillary qubit as
\begin{gather}
    \mathcal{U}_{\mathcal{M}}
    \equiv
    (I_{2^n} \otimes W^{\dagger})
    \cdot
    (
        e^{i\kappa \Theta}\otimes |0\rangle\langle 0|
        +
        e^{-i\kappa \Theta}\otimes |1\rangle\langle 1|
    )
    \nonumber \\
    \cdot
    (I_{2^n} \otimes W\cdot H) .
\end{gather}
This unitary operator transforms the initial state $|\psi\rangle \otimes |0\rangle$ into a superposition of the state which the non-unitary operator $\mathcal{M}$ acts on (success state) and the other state (failure state) as
\begin{gather}
    \mathcal{M}|\psi\rangle \otimes |0\rangle
    +
    \sqrt{1 - \mathcal{M}^2} |\psi\rangle \otimes |0\rangle,
\end{gather}
where an Hermitian operator on $n$ qubits is defined as 
\begin{gather}
    \Theta
    \equiv
    \arccos \frac{\mathcal{M} + \sqrt{1 - \mathcal{M}^2}}{\sqrt{2}},
\end{gather}
the single-qubit unitary operator is
\begin{gather}
    W
    \equiv
    \frac{1}{\sqrt{2}}
    \begin{pmatrix}
        1 & -i \\
        1 & i
    \end{pmatrix} ,
\end{gather}
$\kappa \equiv \mathrm{sgn}(m_0 - 1/ \sqrt{2})$, $H$ denotes the Hadamard gate, and $I_{2^n}$ is the identity operator for an $n$-qubit system.
With probability $P = \langle \psi | \mathcal{M}^2 | \psi\rangle $ the ancillary qubit is observed as a $|0\rangle$ state, then the success state $\mathcal{M}|\psi\rangle$ is obtained.

We consider an approximate implementation since it is difficult to decompose the unitary operator $e^{\pm i\kappa \Theta}$ into a universal gate set for the general Hamiltonian $\hat{\mathcal{H}}$.
The Taylor expansion in the first order of $\Delta\tau$ of the Hermitian operator $\Theta$ is derived as
$
    \kappa \Theta 
    =
    \theta_0 
    - 
    \hat{\mathcal{H}} s_1 \Delta\tau 
    +
    O(\Delta\tau^2), 
$
where $\theta_0 \equiv \kappa \arccos [(m_0 + \sqrt{1-m_0^2})/\sqrt{2}]$ and $s_1 \equiv m_0 / \sqrt{1 - m_0^2}$.
$e^{\pm i\kappa \Theta}$ is approximated by
\begin{gather}
    e^{\pm i\kappa \Theta}
    = 
    e^{\pm i \theta_0}
    e^{\mp i s_1 \Delta\tau \hat{\mathcal{H}}} 
    +
    O(\Delta\tau^2).
\end{gather}
 It is known that the real-time evolution (RTE) operator $e^{\mp i s_1 \Delta\tau \hat{\mathcal{H}}}$ can be implemented in polynomial time for the number of qubit $n$~\cite{Lloyd1996Science, Abrams1997PRL, Kassal2008PNAS, Childs2012QIC, Berry2015PRL, Low2019Quantum, Childs2021PRX}, and thus the quantum circuit for the approximated ITE operator in the first order of $\Delta \tau$ can also be implemented efficiently.
 In this study, we employ a method based on the Suzuki-Trotter decomposition for the implementation of real-time evolution operators~\cite{Lloyd1996Science, Abrams1997PRL, Kassal2008PNAS, Childs2021PRX}.

\subsection{OFDFT using PITE}
\begin{figure*}[ht]
    \centering
    \includegraphics[width=0.85 \textwidth]{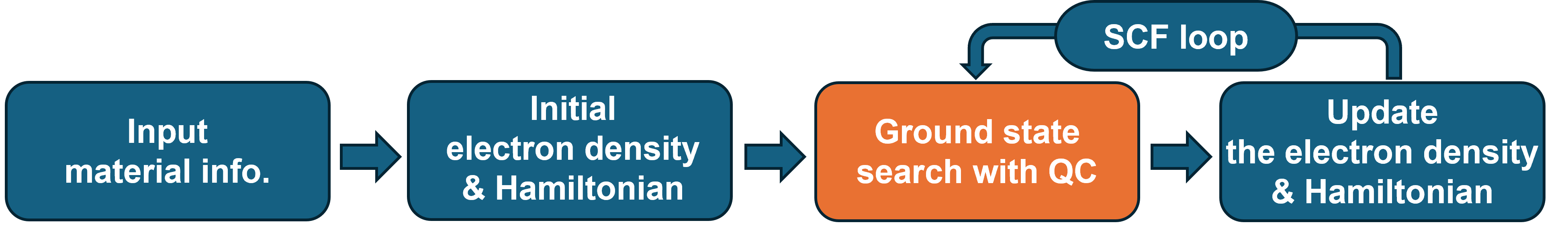}
    \caption{
    Schematic of the execution flow of quantum-classical hybrid scheme of OFDFT calculation. The ground state of the orbital-free Hamiltonian is calculated using quantum computers in the area shown in orange, and the other area in blue is calculated using classical computers.
    }
    \label{fig:flow_QCOFDFT}
\end{figure*}

 An overview of the procedure for quantum-classical hybrid scheme of OFDFT is shown in Fig.~\ref{fig:flow_QCOFDFT}.
 In this study, the first quantized PITE~\cite{Kosugi2022PhysRevResearch} is used for the part of the ground-state calculation of the orbital-free Hamiltonian defined in Eq.~(\ref{eq:of_hamiltonian}), indicated by the orange square.
The other parts, such as the calculation of the next input electron density from the output electron densities and the calculations of $v_\mathrm{r}$ and $v_\mathrm{KS}$, are performed on classical computers.
In this section, we summarize the items necessary to implement the RTE by the orbital-free Hamiltonian, which is required to construct the quantum circuit for PITE. Of course, one can also use the first-quantized adiabatic time evolution~\cite{Nishiya} as a ground state calculation using the implementation of RTE described here.

\subsubsection{Encoding of the electron density}
We encode the square root of the electron density $\phi(\bs{r})\equiv \sqrt{\rho(\bs{r})}$ in a 3-dimensional simulation cell with periodic boundary conditions along the primitive lattice vectors, $\bs{a}_1$, $\bs{a}_2$, and  $\bs{a}_3$.
For each spatial direction $\bs{a}_\ell$, $n_{q\ell}$ qubits are allocated as usual in the first-quantized formalism~\cite{Wiesner1996arxiv, Zalka1998PRSLA, Kassal2008PNAS, Kosugi2022PhysRevResearch, Jones2012NJP, Chan2023sciadv}, and $\bs{a}_\ell$ is divided into equidistant $N_\ell \equiv 2^{n_{q\ell}}$ grid points. 
The spin degree of freedom of the electron density can be encoded by introducing a single additional qubit. We ignore, however, the spin degree of freedom in the present study for simplicity.
 We identify the tensor product of the computational basis in each direction specified by the integers $k_1, k_2, k_3$:
 \begin{align}
    | \boldsymbol{k} \rangle
    \equiv
        | k_1 \rangle_{n_{q 1}}
        \otimes
        | k_2 \rangle_{n_{q 2}}
        \otimes
        | k_3 \rangle_{n_{q 3}}
    \label{QC1q_for_PBC:def_pos_state}
\end{align}
 with the position eigenstate $|\bs{r}\rangle$ at position $\bs{r}=k_1 \bs{a}_1/N_1 + k_2 \bs{a}_2/N_2 + k_3 \bs{a}_3/N_3$.
 The state $|\phi\rangle$ on the quantum register for the square root of the electron density is expressed by amplitude encoding as
 \begin{gather}
|\phi\rangle = 
\sqrt{\frac{V_\mathrm{cell}}{n_eN_\mathrm{g}}}
\sum_{i=0}^{N_\text{g}-1}\phi(\bs{r}_i)|\bs{r}_i\rangle,
\end{gather}
where $N_\text{g}\equiv N_1 N_2 N_3$ is the total number of grid points, $V_\mathrm{cell} \equiv \bs{a}_1\cdot(\bs{a}_2\times\bs{a}_3)$ is the cell volume, and $n_e$ is the number of electrons.
A method has been proposed for such amplitude encoding that can be performed in $O(\mathrm{poly} \log N_\mathrm{g})$ time using Quantum random access memory (QRAM))~\cite{QRAMPhysRevA2008, Prakash:EECS-2014-211}.

\subsubsection{Implementation of the kinetic part}
 Given an electron density $\rho(\bs{r})$, the time evolution operator of the orbital-free Hamiltonian can be written by the Suzuki-trotter expansion as
\begin{gather}
\exp(-i\hat{\mathcal{H}}_\mathrm{OF}\Delta t) = \exp(-i\hat{T}\Delta t)\exp(-i\hat{V}\Delta t) \nonumber \\
+O(\|\hat{T} \| \|\hat{V}\|\Delta t^2),
\end{gather}
where $\hat{T} \equiv  -\nabla^2 /2$, $\hat{V} \equiv v_\mathrm{KS}(\bs{\hat{r}}) + v_\mathrm{r}(\bs{\hat{r}})$. The dependence on the constant $\lambda$ and the electron density $\rho$ is omitted.
Therefore, when $\Delta t$ is chosen small enough, $\exp(-i\hat{T}\Delta t)$ and $\exp(-i\hat{V}\Delta t)$ can be implemented separately and act in sequence.
 Now that we are considering an application to materials calculations, it is reasonable to assume that the grid number $N_\mathrm{g}$ is proportional to the cell volume, number of atoms contained, and number of valence electrons in the system.
 This means, in other words, that when a huge simulation cell is the target of a calculation, e.g., for the simulation of an amorphous system, the same cutoff energy may be set as for the calculation of a small cell.
 Under these assumptions, $\|\hat{T}\|$ and $\|\hat{V}\|$ are independent of $N_\mathrm{g}$. Therefore, $\Delta t$ can also be determined independent of $N_\mathrm{g}$.

This subsection describes how to implement $\exp(-i\hat{T}\Delta t)$ in a periodic system where the simulation cell is defined by primitive lattice vectors $\bs{a}_1, \bs{a}_2, \bs{a}_3$.
Assuming that $\bs{a}_1, \bs{a}_2, \bs{a}_3 $ form a right-handed system and are in general non-orthogonal, the reciprocal primitive lattice vector $\bs{b}_\ell$ is defined as
\begin{align}
    \boldsymbol{b}_1
    \equiv
        \frac{2 \pi}{V_{\mathrm{cell}}}
        \boldsymbol{a}_2 \times \boldsymbol{a}_3
    , \
    \boldsymbol{b}_2
    \equiv
        \frac{2 \pi}{V_{\mathrm{cell}}}
        \boldsymbol{a}_3 \times \boldsymbol{a}_1
    , \
    \boldsymbol{b}_3
    \equiv
        \frac{2 \pi}{V_{\mathrm{cell}}}
        \boldsymbol{a}_1 \times \boldsymbol{a}_2.
\end{align}
 We define discretized momentum in the $\boldsymbol{b}_\ell$ direction, specified by the integer $G_\ell$ as
\begin{align}
    p_\ell^{(G_\ell)}
    \equiv
        G_\ell
        b_\ell
        \
        \left(
            G_\ell
            =
                -\frac{N_\ell}{2},
                -\frac{N_\ell}{2} + 1
                , \dots,
                \frac{N_\ell}{2} - 1
        \right).
\end{align}
Then the momentum eigenstates in the $\boldsymbol{b}_\ell$ direction are defined as
\begin{align}
    | G_\ell \rangle_{\mathrm{mom}}
    &\equiv
        \frac{1}{\sqrt{N}_\ell}
        \sum_{k = 0}^{N_\ell - 1}
        \exp
        \left(
            i
            p_\ell^{(G_\ell)}
            \frac{\boldsymbol{b}_\ell}{|b_\ell|}
            \cdot
            \frac{k}{N_\ell}
            \boldsymbol{a}_\ell
        \right)
        | k \rangle_{n_{q \ell}}
    \nonumber \\
    &=
        \frac{1}{\sqrt{N}_\ell}
        \sum_{k = 0}^{N_\ell - 1}
        \exp
        \left(
            i
            \frac{2 \pi G_\ell k }{N_\ell}
        \right)
        | k \rangle_{n_{q \ell}}.
    \label{QC1q_for_PBC:def_mom_eigenstate}
\end{align}
 Accordingly, we define the operation of the momentum operator $\hat{p}_\ell$ in the $\boldsymbol{b}_\ell$ direction as
\begin{align}
    \hat{p}_\ell
    | G_\ell \rangle_{\mathrm{mom}}
    \equiv
        p_\ell^{(G_\ell)}
        | G_\ell \rangle_{\mathrm{mom}}.
\end{align}
 In this case, the tensor product of the momentum eigenstates in each direction specified by the integers $G_1, G_2, G_3$
\begin{align}
    | \boldsymbol{G} \rangle_{\mathrm{mom}}
    \equiv
        | G_1 \rangle_{\mathrm{mom}}
        \otimes
        | G_2 \rangle_{\mathrm{mom}}
        \otimes
        | G_3 \rangle_{\mathrm{mom}}
    \label{QC1q_for_PBC:def_mom_state_3d}
\end{align}
is the momentum eigenstate in 3-dimensional reciprocal space.
In fact, the momentum operation $\hat{\bs{p}}=\hat{p}_1 \bs{b}_1 / |\bs{b}_1| + \hat{p}_2 \bs{b}_2 / |\bs{b}_2| + \hat{p}_3 \bs{b}_3 / |\bs{b}_3|$ in 3-dimensional reciprocal space on $| \boldsymbol{G} \rangle_{\mathrm{mom}}$ is calculated as
\begin{align}
    \hat{\boldsymbol{p}}
    | \boldsymbol{G} \rangle_{\mathrm{mom}}=
        \left(
            G_1 \boldsymbol{b}_1
            +
            G_2 \boldsymbol{b}_2
            +
            G_3 \boldsymbol{b}_3
        \right)
        | \boldsymbol{G} \rangle_{\mathrm{mom}}.
\end{align}
From Eq.~(\ref{QC1q_for_PBC:def_mom_eigenstate}), the transformation between position and momentum eigenstates in each direction can be written using centered quantum Fourier transform (CQFT)~\cite{Somma2015arxiv, Ollitrault2020PRL} as
\begin{align}
    \mathrm{CQFT}_\ell
    | k \rangle_{n_{q \ell}}
    =
    | k-N_\ell /2 \rangle_{\mathrm{mom}},
    \label{QC1q_for_PBC:action_of_CQFT_on_mom_eigenstate_1d}
\end{align}
where $\mathrm{CQFT}_\ell$ is CQFT operation on $n_{q\ell}$-qubit system.
Therefore, using a diagonal operator in positional basis
\begin{align}
U_\mathrm{kin}(t) \equiv \sum_{\bs{k}} e^{-\frac{it}{2}|\sum_{\ell=1}^3(k_\ell-N_\ell/2)\bs{b_\ell}|^2 } |\bs{k}\rangle\langle \bs{k} |,
\end{align}
the RTE operator by the kinetic energy $\hat{T}=\hat{\bs{p}}^2/2$ is expressed as
\begin{align}
e^{-i \hat{T}t} = 
\sum_{\bs{G}} e^{-\frac{it}{2}|G_1 \boldsymbol{b}_1+G_2 \boldsymbol{b}_2 +G_3 \boldsymbol{b}_3|^2}
|\bs{G}\rangle_\mathrm{mom}\langle \bs{G} |_\mathrm{mom} \nonumber \\
= \mathrm{CQFT}^{(\mathrm{3D})} U_\mathrm{kin}(t) \mathrm{CQFT}^{(\mathrm{3D})\dagger},
\end{align}
where $\mathrm{CQFT^{(3D)}} \equiv \mathrm{CQFT}_1 \otimes \mathrm{CQFT}_2 \otimes \mathrm{CQFT}_3 $.
Now, we introduce the operators acting on the computational basis $|k\rangle_{n_{q\ell}}$ in the $\bs{a}_\ell$ direction or their tensor product defined by
\begin{align}
    U_{\mathrm{kin}, \ell} (t)
    | k \rangle_{n_{q \ell}}
    \equiv
        \exp
        \left(
            -\frac{it}{2}|\bs{b_\ell}|^2
            \left(
                k - \frac{N_\ell}{2}
            \right)^2            
        \right)
        | k \rangle_{n_{q \ell}}
        \label{eq:Ukindiag}
\end{align}
and
\begin{gather}
U_{\mathrm{kin}, \ell \ell'} (t)
    | j \rangle_{n_{q \ell}}
    \otimes
    | k \rangle_{n_{q \ell'}} \nonumber \\
    \equiv
        \exp
        \left(
            -it\bs{b}_\ell \cdot\bs{b}_{\ell'}
            \left(
                j - \frac{N_\ell}{2}
            \right)
            \left(
                k- \frac{N_{\ell'}}{2}
            \right)
        \right)
        | j \rangle_{n_{q \ell}}
        \otimes
        |k \rangle_{n_{q \ell'}}.
    \label{eq:Ukincross}
\end{gather}
Then we can obtain
\begin{align}
U_\mathrm{kin}(t) = \left [ \prod_{\ell=1}^3U_{\mathrm{kin}, \ell}(t) \right]\cdot U_{\mathrm{kin}, 12} (t)U_{\mathrm{kin}, 23} (t)U_{\mathrm{kin}, 31} (t).
\end{align}
If we set $n_{q\ell}=n_q (\ell=1,2,3)$ for the simplicity, the operations in Eqs.~(\ref{eq:Ukindiag}) and ~(\ref{eq:Ukincross}) can be implemented with $O(n_q^2)$ gate numbers as in Ref.~\cite{Kosugi2023JJAP}.
Therefore, the circuit depth of the RTE operator by the kinetic energy is estimated to be $O(\mathrm{poly} \log N_\mathrm{g})$, adding up the circuit depth of the CQFT as well.
Similar to the implementation of the RTE operator by the potential described below, the circuit depth can also be $O(\log N_\mathrm{g})$ by using qubits for redundancy.

\subsubsection{Implementation of the potential part}
\begin{figure*}[ht]
    \centering
    \includegraphics[width=0.85 \textwidth]{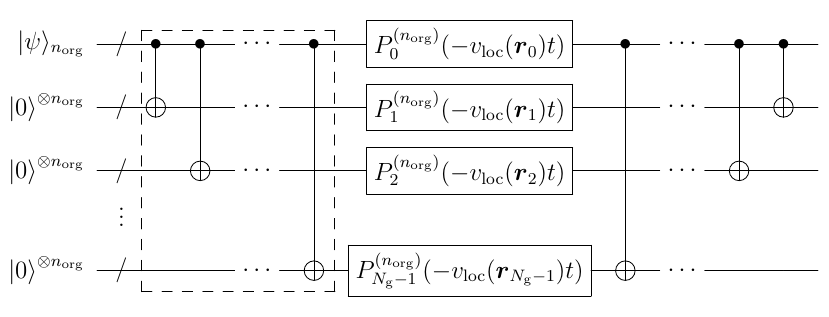}
    \caption{
    Implementation of the real-time evolution by the potential term with circuit depth $O(\log N_g)$ using with redundant registers.
    }
    \label{fig:potential_implement}
\end{figure*}

The real-time evolution by the potential acts diagonally on the positional basis as
\begin{align}
e^{-i\hat{V}t} = \sum_{i = 0}^{N_\mathrm{g}-1}e^{-iv_\mathrm{loc}(\bs{r}_i)t}|\bs{r}_i\rangle\langle\bs{r}_i|,
\end{align}
where $v_\mathrm{loc}(\bs{r})\equiv v_\mathrm{r}(\bs{r}) + v_\mathrm{KS}(\bs{r})$.
This implementation seems to require $O(N_\mathrm{g})$ cost in both number of operations and circuit depth.
However, with the operators defined as
\begin{align}
    P_j^{(n)}(\phi) \equiv I + (e^{i\phi}-1)|j \rangle \langle j| ,
\label{eq:P_j_phi}
\end{align}
the circuit depth can be reduced to $O(\log N_\mathrm{g})$ by adding redundant qubits as shown in Fig.~\ref{fig:potential_implement}, where $n_\mathrm{org}$ is the number of qubits allocated for the electron density register and satisfies $n_\mathrm{org}=O(\log N_\mathrm{g})$.
This reduction in circuit depth is originated from the fact that the sequential CNOT gates in the dashed frame and $P_j^{(n_\mathrm{org})}$ gate can be implemented with $O(\log N_\mathrm{g})$ circuit depth.

\begin{figure}[ht]
    \centering
    \includegraphics[width=0.45 \textwidth]{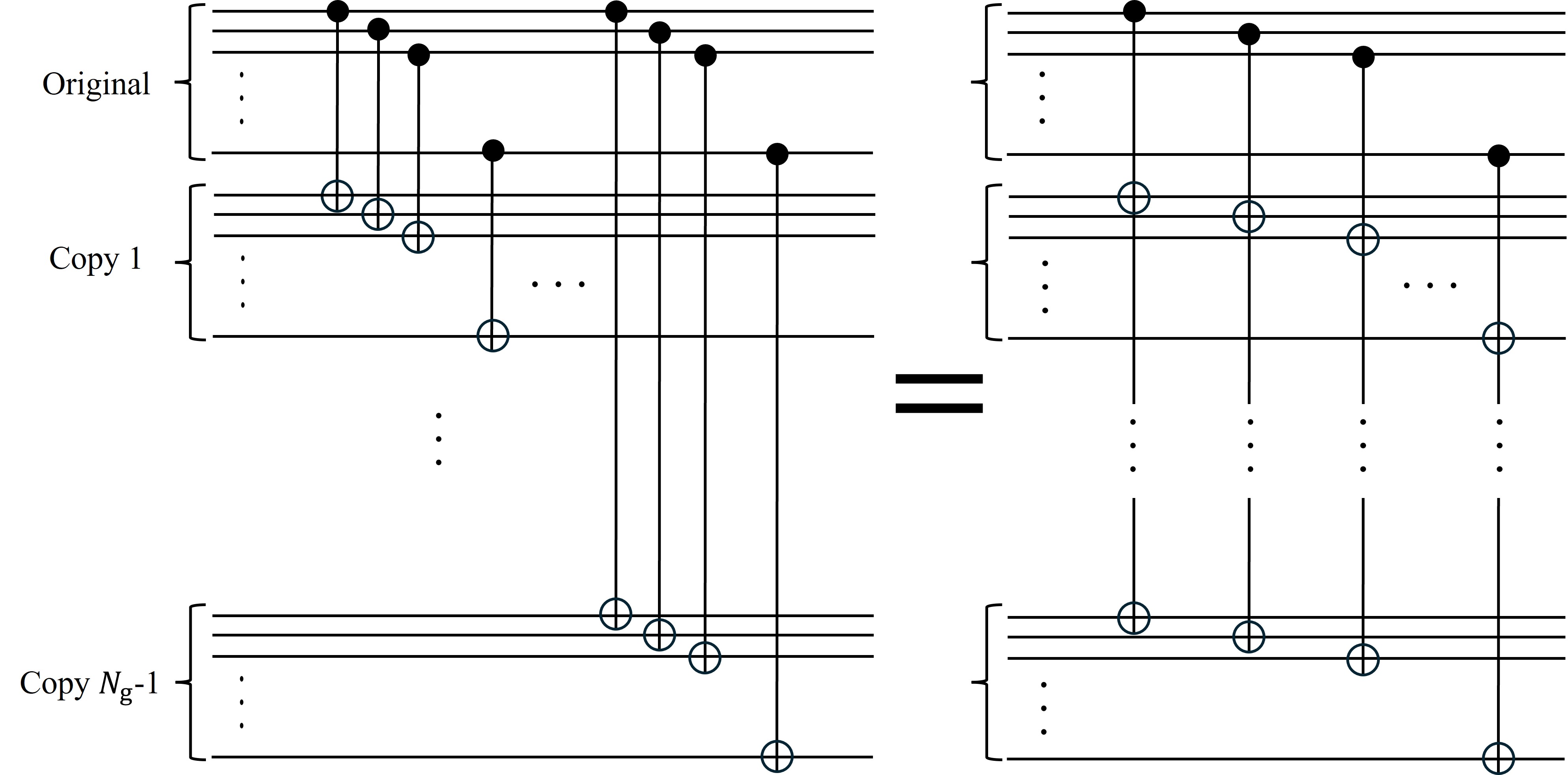}
    \caption{
    Equivalent quantum circuit transformation of the successive CNOT gates shown in the dashed box in Fig.~\ref{fig:potential_implement}.
    }
    \label{fig:fanout}
\end{figure}

First, in the dashed frame shown in Fig.~\ref{fig:potential_implement}, the copy of the computational basis in the register consisting of $n_\mathrm{org}$ qubits is made as
\begin{align}
|j\rangle_{n_\mathrm{org}}\otimes|0\rangle_{n_\mathrm{org}}\otimes \cdots \otimes |0\rangle_{n_\mathrm{org}} \nonumber \\
\rightarrow
|j\rangle_{n_\mathrm{org}}\otimes|j\rangle_{n_\mathrm{org}}\otimes \cdots \otimes |j\rangle_{n_\mathrm{org}}.
\end{align}
As shown in the right-hand side of Fig.~\ref{fig:fanout}, the part in dashed frame is decomposed into $\mathrm{C}X^{\otimes N_\mathrm{g}-1}$ operators on independent qubits, which can be implemented with $O(\log N_\mathrm{g})$ depth by the technique in Ref.~\cite{Gidney2017}.

\begin{figure}[ht]
    \centering
    \includegraphics[width=0.45 \textwidth]{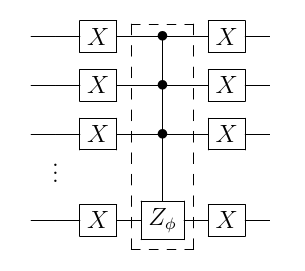}
    \caption{
    Quantum circuit for $P_0^{(n)}(\phi)$, defined in Eq. (\ref{eq:P_j_phi}).
    }
    \label{fig:phaseon0}
\end{figure}

The circuit of $P_0^{(n)}(\phi)$ is then shown in Fig.~\ref{fig:phaseon0}.
For $P_j^{(n)}(\phi)$ we only need to eliminate the two $X$ gates on the qubit corresponding to the digit that is 1 in the binary representation of $j$.
The area enclosed by the dashed line is known to be implemented with a circuit depth of $O(n)$ by adding one ancillary qubit~\cite{Barenco1995PRA}.

Eventually, the circuit depth of the potential part is $O(\log N_\mathrm{g})$.
For the implementation of PITE, a time evolution operation by the potential controlled by the ancillary qubit is required, which can be implemented in a similar manner with a depth of $O(\log N_\mathrm{g})$ by making a copy of the register that also includes the ancillary qubit for PITE, as in Fig.~\ref{fig:potential_implement}.

\section{Application}
\begin{figure*}[ht]
    \centering
    \includegraphics[width=0.85 \textwidth]{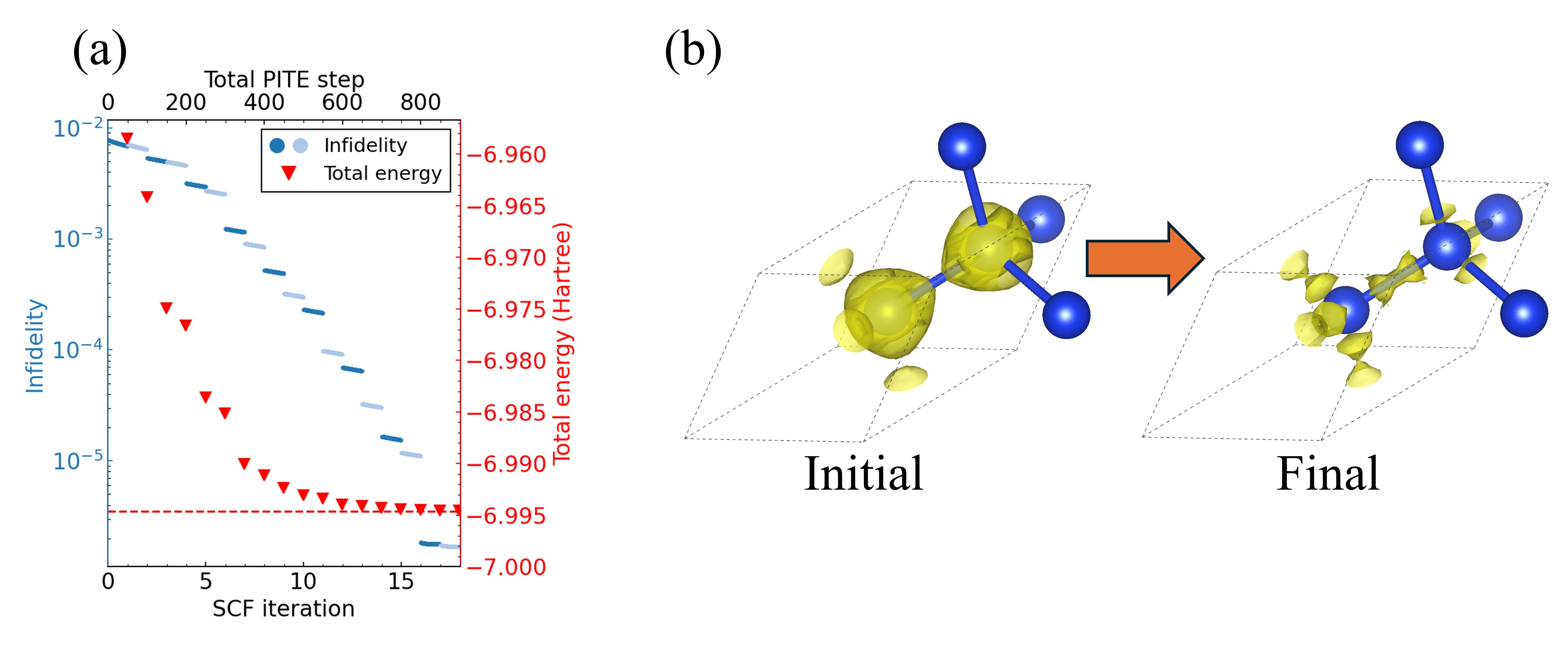}
    \caption{
    Results of the ground-state calculation for the Si primitive cell obtained by OFDFT with first-quantized PITE.
    (a) Blue lines indicate the infidelity defined in Eq.~(\ref{eq:infidelity}). Odd SCF iterations are marked with dark blue lines and even SCF iterations with light blue lines. The red lower triangle indicates the total energy in each SCF iteration. The red dotted line is the converged value of the total energy.
    (b) Electron density in the initial state and the electron density obtained by this method are represented by yellow isosurfaces.
    The isosurfaces are displayed at 90 \% of the maximum value.
    }
    \label{fig:Si2result}
\end{figure*}

The ground state electron density calculation of the Si primitive cell was simulated by OFDFT using PITE in first-quantized formalism.
LKT (Luo-Karasiev-Trickey)~\cite{LKT} was selected for the kinetic energy functional and PBE (Perdew-Burke-Ernzerhof)~\cite{PBE} for the exchange correlation functional, both implemented in LIBXC~\cite{LIBXC} were used.
LIPS in Ref.~\cite{Imoto2021PRR} was used to represent the interaction between the valence electrons and nucleus with core electrons in the OFDFT calculation.
The spatial grid for the representation of the electron density and potential was set to $16\times16\times16$ points and an arbitrary constant $\lambda$ to 1.0.
In this simulation, $U_\mathrm{kin}(t)$ and $\exp (-iv_\mathrm{loc}t)$ were implemented as diagonal matrices and a total of 13 qubits were used, including the ancillary qubit of PITE.
The initial electron density of the system was set to the sum of the electron densities calculated by Kohn-Sham DFT with the GTH-type pseudopotential~\cite{GTH1996} in advance for an isolated Si single atom.
If the input electron density at the $k$-th SCF iteration is $\rho_k^\mathrm{in}(\bs{r})$, $|\phi_k^{\mathrm{gs}}\rangle$ denotes the ground state of the orbital-free Hamiltonian $\hat{\mathcal{H}}_\mathrm{OF}[\rho_k^\mathrm{in}]$.
At this SCF iteration, $|\phi_{k,j}\rangle$ denotes the state after performing $j$ PITE steps, and infideilty is defined as
\begin{gather}
\text{Infidelity} = 1-|\langle \phi_{k,j}|\phi_k^{\mathrm{gs}}\rangle|^2,
\label{eq:infidelity}
\end{gather}
where each SCF iteration starts from $\langle\bs{r}|\phi_{k,0}\rangle = \sqrt{\rho_k^\mathrm{in}(\bs{r})}$.

The behaviour of infidelity and the total energy in DFT when 50 PITE steps are executed per SCF iteration with the PITE time step set to $\Delta t = 0.001$ and the parameter $m_0 = 0.99$ are shown in Fig.~\ref{fig:Si2result}(a).
This simulation was performed under the assumption that the PITE output state $|\phi_{k,50}\rangle$ can be read out by quantum state tomography~\cite{Nielsen2000Book} or other methods and that the output electron density $\rho_k^\mathrm{out}(\bs{r})\equiv|\langle\bs{r}|\phi_{k,50}\rangle|^2$ can be obtained.
The Broyden method~\cite{Broyden1965} was used to update the electron density.
The total success probability of PITE in the $j$PITE steps is given as $P_{j} = m_0^{j} |\langle \phi_{k,j}|\phi_{k}^{\mathrm{gs}}\rangle|^2$ ~\cite{Kosugi2022PhysRevResearch, Nishi2023PRRes}, and the total success probability of PITE in each SCF obtained by numerical calculation was $P_{50} = 0.595 \pm 0.001$.
We observed a behavior in which the total energy decreases with each SCF step update, although the improvement in infidelity due to the 50-step PITE was slight. This result indicates that the small infidelity improvement of PITE plays an important role in the convergence of OFDFT.
With this scheme, the total energy converges in the range of $1\times10^{-5}$ Hatree after 18 SCF iterations, and the electron density changes from the sum of isolated atoms to $sp^3$ bonding state of Si, as shown in Fig.~\ref{fig:Si2result}(b).

\section{Discussions}
\begin{figure}[ht]
    \centering
    \includegraphics[width=0.45 \textwidth]{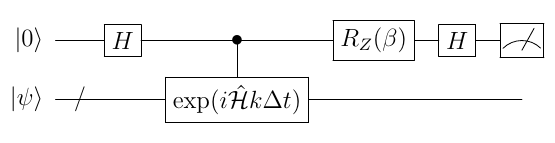}
    \caption{
    Quantum circuit for Baysian QPE
    }
    \label{fig:QPEcirc}
\end{figure}

We have mentioned that ground-state calculations based on RTE operators such as PITE can be performed in $O(\log N_\mathrm{g})$ time, however for updating the electron density and OFDFT Hamiltonian, the state of the quantum registers, which are the output of such ground-state calculations, must be obtained as classical information.
To determine the probability amplitude of each computational basis, i.e., the electron density on the corresponding grid point, by direct observation of the register, $O(N_\mathrm{g})$ times observation is required.
The computational cost of OFDFT using the classical algorithm is reported to be $O(N_\mathrm{g})$ to $O(N_\mathrm{g}\log N_\mathrm{g})$, which means the number of direct observations of the register is comparable to the classical computational time.
One way to avoid this is to use QPE~\cite{Kitaev1995arXiv, Abrams1999PRL, Wiebe2016PRL, OBrien_2019NJP, Yamamoto2024PRR, Ding2023PRXQ} to efficiently obtain only the minimum eigenvalue.
After preparing the quantum register state representing the ground state of the Hamiltonian $\hat{\mathcal{H}}$ by PITE, the circuit shown in Fig.~\ref{fig:QPEcirc} is repeatedly performed with an ancillary qubit initialized to $|0\rangle$.
This is an example of Bayesian QPE~\cite{Wiebe2016PRL, OBrien_2019NJP, Yamamoto2024PRR}. The probability of obtaining $m\in\{{0,1\}}$ by observation of ancillary qubit when inputting the eigenstate corresponding to eigenvalue $\mu$ of $\hat{\mathcal{H}}$ is given by
\begin{align}
    p(m|\mu,k,\beta) = \frac{1+\cos (k\mu\Delta t + \beta -m\pi)}{2},
\end{align}
and the distribution of eigenvalues is estimated based on Bayesian estimation by obtaining observation results of the ancillary qubit while updating $k,\beta$.
Since the QPE circuit also requires RTE operator with a given Hamiltonian, the circuit depth required to estimate the minimum eigenvalue within a given accuracy can be evaluated from the above discussion as $O(\log N_\mathrm{g})$.
One example of utilizing the minimum eigenvalue obtained and efficiently acquiring its eigenvector by a classical algorithm is to employ an iterative eigensolver, such as the locally optimal block preconditioned conjugate gradient (LOBPCG) method, to prepare the appropriate preconditioner and reduce the number of iterations.
In the LOBPCG method, the computational cost per iteration is determined by the matrix-vector product, so it can be performed with a cost of $O(N_\mathrm{g})$ for a sparse Hamiltonian.
When the minimum eigenvalue $\mu_\mathrm{gs}$ of the Hamiltonian $\hat{\mathcal{H}}$ is known, 
the preconditioner defined as
\begin{align}
    M = (\hat{\mathcal{H}}-\mu_\mathrm{gs}I)^{-1}
\end{align}
leads to rapid convergence to the eigenstate corresponding to $\mu_\mathrm{gs}$.
Although exact inverse matrix computation requires a computational cost of $O(N_\mathrm{g}^2)$ even for sparse matrices, the user's choice of approximate inverse matrix computation methods such as the Point-Jacobi method or incomplete LU factorization may reduce the overall computation time.

\section{Conclusions}
In this study, we proposed a quantum-classical hybrid scheme of orbital-free density functional theory (OFDFT) as a first step towards a materials calculation method for large-scale systems using quantum computer.
Specifically, we adopted OFDFT, which repeatedly determines the ground state of the single-electron Schrödinger equation, and performed the first quantized form of the probabilistic imaginary-time evolution (PITE) and quantum phase estimation (QPE) for the ground state calculation and smallest eigenvalue estimation, respectively, with circuit depth $O(\log N_\mathrm{g })$.
Indeed, the electron density of the ground state composed of $sp^3$ bonds of the Si primitive cell is successfully obtained by updating the electron density using PITE and the Broyden method.

Further studies would include a more efficient implementation of real-time evolution (RTE) operator generated by potential terms and a more efficient readout of the output state of PITE sotored in quantum register.
For realistic material calculation problem sizes and computational objectives, it is also important to know how much benefit is provided by the use of the QPE which gives the lowest eigenvalue to obtain the electron density of the ground state.
As a further application of quantum computation, it may also be useful to apply it to updating the orbital-free Hamiltonian, i.e. the part where $v_\mathrm{loc}(\bs{r})$ is calculated from the electron density.
Another approach to large scale materials calculations has been proposed to perform the Kohn-Sham DFT (KSDFT)~\cite{KohnSham1965}, which is also a quantum-classical hybrid scheme~\cite{ko2023arxiv}.
Although it is difficult to calculate the total energy using this method because the Kohn-Sham orbitals are not obtained directly, it is worth noting that KSDFT, which is said to have a computational cost of $O(N_\mathrm{g}^3)$ on classical computers, can be performed in $O(N_\mathrm{g})$ using quantum computers.

\begin{acknowledgments}
The author acknowledges the contributions and discussions provided by the members of Quemix Inc.
The authors thank the Supercomputer Center, the Institute for Solid State Physics, the University of Tokyo for the use of the facilities.
This work was supported by JSPS KAKENHI under Grantin-Aid for Scientific Research No.21H04553, No.20H00340, and No.22H01517, JSPS KAKENHI. This work was partially supported by the Center of Innovations for Sustainable Quantum AI (JST Grant Number JPMJPF2221).

\end{acknowledgments}


\bibliographystyle{apsrev4-2}
\bibliography{ref}

\end{document}